\documentclass[acmtog,nonacm]{acmart}

\acmSubmissionID{262}

\usepackage{booktabs} 

\usepackage{diagbox}

\usepackage[ruled]{algorithm2e} 

\SetAlFnt{\small}
\SetAlCapFnt{\small}
\SetAlCapNameFnt{\small}
\SetAlCapHSkip{0pt}

\acmJournal{TOG}

\usepackage[normalem]{ulem}

\usepackage{caption}
\usepackage{subcaption}

\begin{document}



\title{Learning Audio-Driven Viseme Dynamics for 3D Face Animation}



\author{Linchao Bao}
\affiliation{%
  \institution{Tencent AI Lab}
  \city{Shenzhen}
  \country{China}
}
\email{linchaobao@gmail.com}

\author{Haoxian Zhang}
\affiliation{%
  \institution{Tencent AI Lab}
  \city{Shenzhen}
  \country{China}
}

\author{Yue Qian}
\affiliation{%
  \institution{Tencent AI Lab}
  \city{Shenzhen}
  \country{China}
}

\author{Tangli Xue}
\affiliation{%
  \institution{Tencent AI Lab}
  \city{Shenzhen}
  \country{China}
}

\author{Changhai Chen}
\affiliation{%
  \institution{Tencent AI Lab}
  \city{Shenzhen}
  \country{China}
}

\author{Xuefei Zhe}
\affiliation{%
  \institution{Tencent AI Lab}
  \city{Shenzhen}
  \country{China}
}

\author{Di Kang}
\affiliation{%
  \institution{Tencent AI Lab}
  \city{Shenzhen}
  \country{China}
}


\begin{abstract}
We present a novel audio-driven facial animation approach that can generate realistic lip-synchronized 3D facial animations from the input audio. 
Our approach learns viseme dynamics from speech videos, produces animator-friendly viseme curves, and supports multilingual speech inputs. 
%
%
The core of our approach is a novel parametric viseme fitting algorithm that utilizes phoneme priors to extract viseme parameters from speech videos. 
%
With the guidance of phonemes, the extracted viseme curves can better correlate with phonemes, thus more controllable and friendly to animators. 
To support multilingual speech inputs and generalizability to unseen voices, we take advantage of deep audio feature models pretrained on multiple languages to learn the mapping from audio to viseme curves. 
%
Our audio-to-curves mapping achieves state-of-the-art performance even when the input audio suffers from distortions of volume, pitch, speed, or noise. 
Lastly, a viseme scanning approach for acquiring high-fidelity viseme assets is presented for efficient speech animation production. 
We show that the predicted viseme curves can be applied to different viseme-rigged characters to yield various personalized animations with realistic and natural facial motions. 
Our approach is artist-friendly and can be easily integrated into typical animation production workflows including blendshape or bone based animation. 
\end{abstract}

\ccsdesc[500]{Computing methodologies~Animation}
\ccsdesc[500]{Computing methodologies~Neural networks}

\keywords{3D facial animation, digital human, lip-sync, viseme, phoneme, speech, neural network}

\begin{teaserfigure}
  \includegraphics[width=\textwidth]{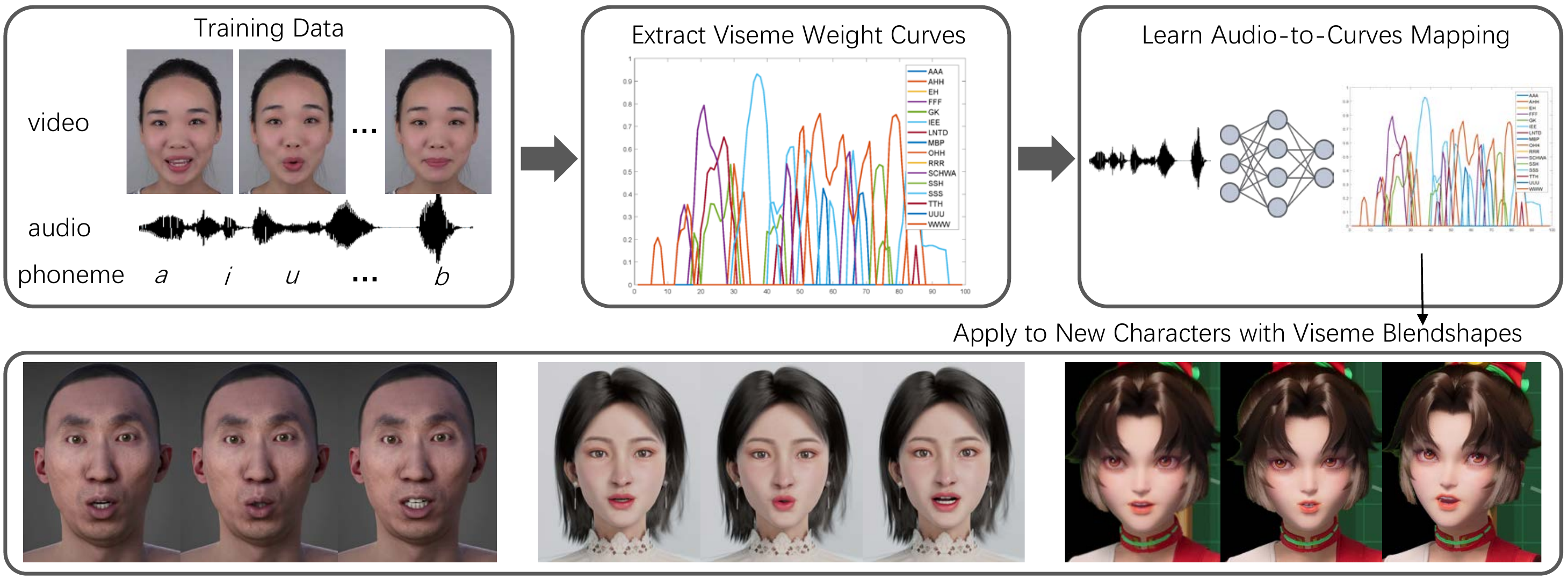}
  \vspace{-5mm}
  \caption{Overview of our audio-driven facial animation approach. 
  We present a novel phoneme-guided facial tracking algorithm to extract viseme curves from videos. An audio-to-curves mapping model is then learned from the paired audio and viseme curve data. The learned model can be used to produce speech animation from audio for characters in various styles.}
  \Description{The pipeline illustration of our approach.}
  \vspace{3mm}
  \label{fig:teaser}
\end{teaserfigure}

\maketitle

\section{Introduction}

Realistic facial animation for digital characters is essential in various immersive applications like video games and movies. Traditionally, creating facial animation that is synchronized to spoken audio is very time-consuming and challenging for artists. Facial performance capture systems were developed to ease the difficulties. However, animation production with performance capture systems is still challenging in many cases due to the required 
professional workflows and actors. There is a vast demand to investigate technologies that can conveniently produce high-fidelity facial animations with solely audio as input. 

Recent advances in speech-driven facial animation technologies can be categorized into two lines: vertex-based animation \cite{karras2017audio,cudeiro2019capture,richard2021meshtalk,liu2021geometry,fan2022faceformer} and parameter-based animation \cite{edwards2016jali,taylor2017deep,zhou2018visemenet,huang2020speaker}. Vertex-based animation approaches directly learn mappings from audio to sequences of 3D face models by predicting mesh vertex positions, while parameter-based animation approaches instead generate animation curves (i.e., sequences of animation parameters). Though the former approaches draw increasing attention in recent years, the latter approaches are more suitable for animation production as their results are more controllable and can be integrated into typical animator workflows more conveniently. 

The parameter space of animation curves can be defined by either an algorithm-centric analysis like Principal Component Analysis (PCA) \cite{taylor2017deep,huang2020speaker}, or an animator-centric rig defined by artists like JALI \cite{edwards2016jali,zhou2018visemenet}. The algorithm-centric parameter space is easy to derive from facial performance data but is usually difficult to be applied to new characters. The animator-centric parameter space relies on specific facial rigs crafted by artists and is usually more friendly to animators. JALI \cite{edwards2016jali} defines a state-of-the-art facial rig based on viseme shapes that can be varied by independently controlling jaw and lip actions. However, the flexible yet complicated rigging makes it difficult to utilize facial performance data to learn the animation. Although an elaborated procedural animation approach is proposed \cite{edwards2016jali} to produce expressive facial animations 
from audio, the artificially generated dynamics are unnatural. The deep learning approach VisemeNet \cite{zhou2018visemenet} which is trained from procedurally generated data also has the same problem.

In this paper, we propose a novel approach that employs an artist-friendly parameter space for animation and learns the animation curves from facial performance data. Specifically, each frame of our animation curves consists of a set of weights for viseme blendshapes. We first extract the viseme weights from facial performance videos using a novel phoneme-guided facial tracking algorithm. Then the mapping from audio to viseme curves is learned using neural networks. Since each viseme has clear meanings, the preparation of viseme blendshapes for new characters is very straightforward and artist-friendly. It can be achieved by scanning the visemes of human actors or manually crafting personalized visemes for stylized characters by artists. As the viseme weight curves extracted from videos represent the dynamics of facial motions, the final animations by applying the curves are more natural than those that are procedurally generated. Most importantly, the guidance of the phonemes during facial tracking reduces the ambiguity of the parametric fitting process, and thus the resulting viseme weights are more correctly correlated with phonemes. This makes the animation curves more friendly to animators, and more importantly, largely improves the generalization performance when applied to new characters.

The contributions of this paper include: 
\begin{itemize}
  \item A novel phoneme-guided 3D facial tracking algorithm for speech-driven facial animation, which produces more correctly activated viseme weights and is  artist-friendly. 
  \item A state-of-the-art neural network model for mapping audio to animation curves, which supports multilingual speech inputs and generalizes well to unseen speakers.
  \item An efficient procedure of high-fidelity digital human animation production by scanning the visemes of human actors.
\end{itemize}

\section{Related Work}

\subsubsection*{Facial Performance Capture from Video}

Facial performance capture from RGB videos has been an active research area for many years \cite{beeler2011high,garrido2013reconstructing,shi2014automatic,cao2015real,thies2016face2face,cao2018stabilized,laine2017production,zoss2019accurate,zhang2022video}. A full review of this area is beyond the scope of this paper (please refer to the surveys \cite{zollhofer2018state,egger20193d}). We mainly focus on methods that can extract facial expression parameters from videos. Existing methods typically achieve this by fitting a linear combination of blendshapes \cite{weise2011realtime,garrido2013reconstructing} or multi-linear morphable models \cite{shi2014automatic,thies2016face2face} to each frame of an input video, by optimizing objective functions considering feature alignment, photo consistency, statistical regularization, etc. Besides, learning based methods \cite{cao20133d,cao2015real} are also developed to further improve the inference efficiency. The key difference between blendshapes \cite{lewis2014blendshape} and multi-linear models \cite{cao2014facewarehouse} is that the space of a set of blendshape weights is usually defined by artists and not necessarily orthogonal, while a multi-linear model is typically derived by Principal Component Analysis (PCA) and the parameter space is orthogonal. Thus blendshape fitting with different priors might lead to different solutions. In this paper, we employ a viseme-blendshape space and introduce a novel phoneme guided prior to the fitting.

\subsubsection*{Parameter-Based 3D Facial Animation from Audio}

Viseme curves based animation from audio dates back to the work by Kalberer et al. \shortcite{kalberer2002speech} in early 2000's. They construct a viseme space through Independent Component Analysis (ICA) and fitting splines to sparse phoneme-to-viseme mapped points within the space to produce smooth animation. However, the splines cannot well represent the real dynamics of facial motions and thus the animation is unnatural. Taylor et al. \shortcite{taylor2012dynamic} proposed an approach to discover temporal units of dynamic visemes to incorporate realistic motions into the motion units. The final animation is generated by simply concatenating the units, which would still introduce unnatural transitions between the units. Their subsequent work \cite{taylor2017deep} based on deep learning improved over this problem by directly learning natural coarticulation motions from data, but it relies on an Active Appearance Model (AAM) based parameter space, which is not artist-friendly and would degrade the animation quality when retargeted to new characters. Edwards et al. \shortcite{edwards2016jali} developed a complicated procedural animation approach based on a viseme-based rig with independently controllable jaw and lip actions, namely JALI. The animation curves are generated based on explicit rules with empirical parameters. Though JALI is widely adopted in game industries due to its artist-friendly design and versatile capability, the generated facial movements are with a sense of artificial composition. The problem still exists in the subsequent work VisemeNet \cite{zhou2018visemenet}, which employs a deep learning approach to learn the animation curves from data generated by JALI.

\subsubsection*{Vertex-Based 3D Facial Animation from Audio}

Another line of the work aims to directly produce vertex-based animation from audio \cite{karras2017audio,cudeiro2019capture,richard2021meshtalk,liu2021geometry,fan2022faceformer}, mainly based on deep learning approaches. Karras et al. \shortcite{karras2017audio} proposed a neural network to directly learn a mapping from audio to vertex positions of a face model, based on 3-5 minutes high-quality tracked mesh sequences of a specific character. The approach is subject-specific and thus animating a new character requires either new high-cost training data or a deformation retargeting step that lacks personalized control. VOCA \cite{cudeiro2019capture} models multi-subject vertex animation with a one-hot identity encoding, which is trained from a 12-speaker 4D scan dataset. MeshTalk \cite{richard2021meshtalk} further extends the work from lower-face to whole-face animation and employs a larger dataset containing 4D scans of 250 subjects. However, these vertex-based animation approaches are not artist-friendly and are difficult to be integrated into animation production workflows. On the other hand, the neural network models in these approaches including Temporal Convolutional Networks (TCNs) \cite{karras2017audio,cudeiro2019capture,liu2021geometry,richard2021meshtalk}, Long Short-Term Memory Networks (LSTMs) \cite{zhou2018visemenet,huang2020speaker}, Transformers \cite{fan2022faceformer}, etc., provide us valuable references in developing our audio-to-curves mapping model.







\section{Overview}

We first describe the dataset and visemes used in our approach. Then we briefly summarize the pipeline of our approach and the organization of this paper. 

\subsubsection*{Dataset}

We collected a dataset containing 16-hour speaking videos of a Chinese actress. The dataset has roughly $12,000$ utterances ($10,000$ for training and $2,000$ for testing), mostly in Chinese with a small quantity containing English words. It is  synchronously recorded by a video camera and a high-fidelity microphone. Text transcripts for the audio are annotated by humans. Phonemes with start/end timestamps are generated with an in-house phonetic alignment tool similar to Montreal Forced Aligner \cite{mcauliffe2017montreal}.

\subsubsection*{Visemes and Phonemes}

\begin{figure}[t!]
  \centering
  \includegraphics[width=0.98\linewidth]{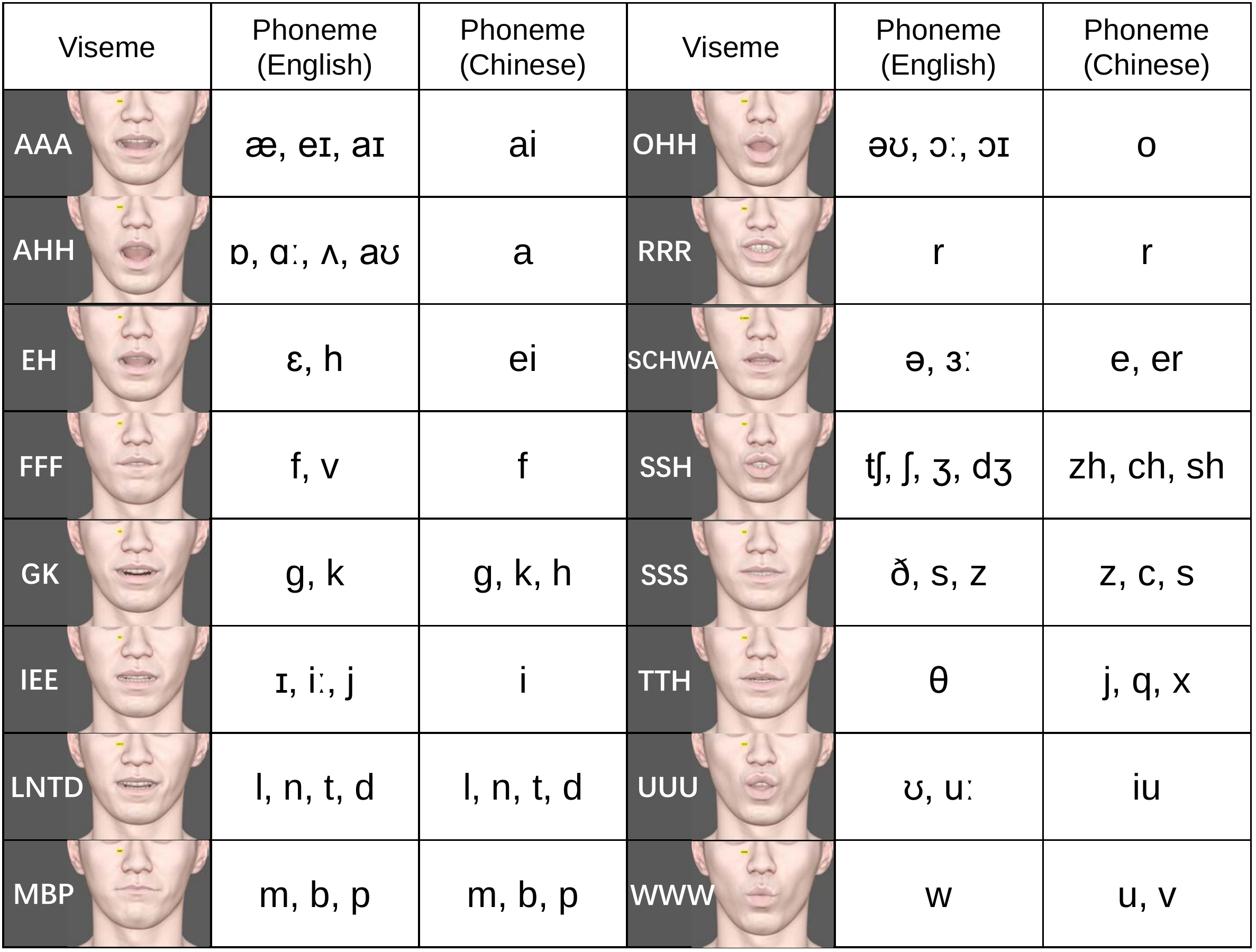}
  \vspace{-2mm}
  \caption{The 16 visemes and their corresponding phonemes in our approach. The English phonemes are in International Phonetic Alphabet (IPA) notation, and the Chinese phonemes are in Chinese Pinyin notation.}
  \Description{The visemes and their corresponding phonemes used in this paper.}
  \label{fig.viseme16}
\end{figure}

We use a set of 16 visemes as shown in Figure \ref{fig.viseme16}, with minor differences from JALI \cite{edwards2016jali} and Visemenet \cite{zhou2018visemenet}. The corresponding English and Chinese phonemes are also shown in the figure. Note that the phoneme labels for the audio in our dataset consist of both English and Chinese phonemes, which are mapped to the same set of visemes.

\subsubsection*{Approach}

In our approach, we first extract viseme weights for each frame in the videos using a novel phoneme-guided facial tracking algorithm detailed in Section \ref{sec.tracking}. 
Then an audio-to-curves neural network model is trained using the viseme weight curves and their corresponding audio tracks, which is described in Section \ref{sec.mapping}. 
We further present an approach to efficiently acquire high-fidelity viseme assets in Section \ref{sec.animproduction} for applying the audio-driven animation production to new characters. 
The supplementary video can be found on our project page\footnote{\url{https://linchaobao.github.io/viseme2023/}}.


\section{Phoneme-Guided 3D Facial Tracking}\label{sec.tracking}

In this section, we introduce a novel algorithm to extract viseme weights from videos with phoneme labels. The goal of our algorithm is to fit a linear combination of viseme blendshapes for each video frame. Specifically, assume the neutral 3D model of a subject is $B_0$, the set of viseme blendshapes for the subject is $\mathcal{B} = \{ B_1, B_2, ..., B_{16} \}$, where each blendshape $B_i$ represents the 3D model of a viseme shape. The fitting process is to find for each frame $j$ a set of viseme weights $\mathcal{X}^j = \{ x_i ~|~ i \in \{1,2,...,16\}, 0 \leq x_i \leq 1 \}$, such that $S = B_0 + \sum_{i=1}^{16}x_i (B_i - B_0)$ resembles the 3D shape of the subject in each video frame. In order to solve the problem with the guidance of phonetic information, we first present two preparation steps before describing the fitting algorithm. 

\subsection{Procedural Viseme Weights Generation}\label{sec.procvism}

The first step is to procedurally generate viseme weights for each frame of the audio tracks with the phoneme labels. We use a similar algorithm to JALI \cite{edwards2016jali} to generate the viseme weight curves for all audio tracks in our dataset. The curves are generated based on explicit rules regarding the durations of each phoneme's onset, sustain, and offset, with considerations of the phoneme types and co-articulations. Although the viseme weights for frames at phoneme apexes are accurate, the motion dynamics of the artificially composited curves are rather different from real-world motions, which would lead to unnatural animations especially for realistic characters. Assume the generated viseme weights for each frame $j$ are denoted as $\mathcal{A}^j = \{ a_i ~|~ i \in \{1,2,...,16\}, 0 \leq a_i \leq 1 \ \}$.

\subsection{Viseme Blendshapes Preparation}

The second step is to construct the neutral 3D face model and 16 viseme blendshapes for the actress in our dataset. While this step can be largely eased with a camera array device as described in Section \ref{sec.visemescan}, we only have access to the video data of the actress in our large-corpus dataset. Thus we resort to an algorithmic approach to prepare the 3D models. First, we manually select a few video frames of the neutral face of the actress, and reconstruct her 3D face model and texture map using a state-of-the-art 3DMM fitting algorithm \cite{chai2022realy}. Then we transfer the mesh deformation from a set of template viseme models created by artists to the actress's face model using a deformation transfer algorithm \cite{sumner2004deformation}. For each viseme model, we further use the frames at the corresponding phoneme apexes to optimize the viseme shapes with Laplacian constraints \cite{sorkine2004laplacian}. A set of personalized viseme blendshapes of the actress is finally obtained.

\subsection{Phoneme-Guided Viseme Parametric Fitting} \label{sec.vism_fitting}

Our fitting algorithm is based on the 3DMM parametric optimization framework of HiFi3DFace \cite{bao2021high}, with several critical adaptations. First, we iterate through the frames of each video clip to optimize the parameters, taking into consideration of optical flow constraints and temporal consistency. Second, we introduce new loss functions regarding phoneme guidance. Specifically, the loss function to be minimized for each frame $j$ is defined as: 
\begin{multline}\label{eq.totaloptloss}
    L(\mathcal{X}^{j}) = w_{1}L_{lmk} + w_{2}L_{rgb} 
    + w_{3}L_{sup}  + w_{4}L_{act} \\
    + w_{5}L_{flow} + w_{6}L_{diff} + w_{7}L_{range},
\end{multline}
where each term is defined as follows.

\subsubsection*{Landmark Loss $L_{lmk}$ and RGB Photo Loss $L_{rgb}$}

These two losses are similar to HiFi3DFace \cite{bao2021high}. The landmark loss is   
\begin{equation}
    L_{lmk}(\mathcal{X}^{j}) = \frac{1}{ | \mathcal{P} | } \sum_{k \in \mathcal{P}} \beta_k || \mathrm{Proj}(\mathbf{s}_k) - \mathbf{p}_k ||^2_2,
\end{equation}
where $k \in \mathcal{P}$ is a detected landmark point, $\mathrm{Proj}(\cdot)$ is the camera projection of a 3D vertex $\mathbf{s}_k$ to a 2D position, $\mathbf{p}_k$ is the 2D position of a landmark point $k$, and $\beta_k$ is the weight to control the importance of each point. The RGB photo loss is defined as 
\begin{equation}
    L_{rgb}(\mathcal{X}^{j}) = \frac{1}{ | \mathcal{F} | } \sum_{p \in \mathcal{F}} || I_p - I'_p ||^2_2,
\end{equation}
where $\mathcal{F}$ is the set of pixels within facial regions, $I_p$ is the RGB pixel values at pixel $p$, and $I'_p$ is the rendered pixel values with a differentiable renderer \cite{genova2018unsupervised}. All pixel values are normalized to the range of $[0,1]$.


\subsubsection*{Phoneme-Guided Suppression Loss $L_{sup}$ and Activation Loss $L_{act}$}

We use the procedurally generated viseme curves from phoneme labels in Section \ref{sec.procvism} to suppress and activate corresponding visemes. Note that the motion dynamics of the procedurally generated curves are not desired, but the visemes that need to be suppressed or activated in each frame are accurately depicted in the curves. To exploit the information, we first determine two sets of viseme indices $\mathcal{I}^j_{sup}$ and $\mathcal{I}^j_{act}$ for each frame $j$ as follows: $\mathcal{I}^j_{sup}$ consists of viseme indices $i$ such that $a_i$ does not appear in the top $m$ largest values in all neighboring frames of $j$, and $\mathcal{I}^j_{act}$ consists of viseme indices $i$ such that $a_i$ appears in the top $n$ largest values in current frame $j$. We use $m=3$ and $n=2$ in our approach. The idea is that if a viseme appears significant in the current frame, it should be activated. On the other hand, if a viseme does not appear significant in all neighboring frames, it should be suppressed. The corresponding loss terms are defined as
\begin{equation}
    L_{sup}(\mathcal{X}^{j}) =  \frac{1}{ | \mathcal{I}^j_{sup} | } \sum_{i \in \mathcal{I}^j_{sup}} x^2_i,
\end{equation}
\begin{equation}
    L_{act}(\mathcal{X}^{j}) = - \frac{1}{ | \mathcal{I}^j_{act} | } \sum_{i \in \mathcal{I}^j_{act}} x^2_i.
\end{equation}

\subsubsection*{Optical Flow Loss $L_{flow}$ and Temporal Consistency Loss $L_{diff}$}

For each frame $j$, we compute the bi-directional optical flow between frame $j-1$ and $j$ and screen valid correspondences using a forward-backward consistency check \cite{bao2014fast}. The optical flow loss $L_{flow}$ is defined as 
\begin{equation}
    L_{flow}(\mathcal{X}^{j}) = \frac{1}{ | \mathcal{K} | } \sum_{k \in \mathcal{K} } || \mathrm{Proj}(\mathbf{s}^j_k) - (\mathrm{Proj}(\mathbf{s}^{j-1}_k) + \mathbf{u}_k) ||^2_2 , 
\end{equation}
where $k \in \mathcal{K}$ is a valid vertex correspondence from frame $j-1$ to $j$, $\mathrm{Proj}(\cdot)$ is the camera projection of a 3D vertex $\mathbf{s}_k$ to a 2D position, and $\mathbf{u}_k$ is the 2D optical flow displacement for the vertex. Furthermore, an explicit temporal consistency loss $L_{diff}$ over the viseme weights is defined as
\begin{equation}
    L_{diff}(\mathcal{X}^{j}) = \frac{1}{ 16 } \sum_{i=1}^{16} (x_i^j - x_i^{j-1})^2 . 
\end{equation}

\subsubsection*{Value Range Loss $L_{range}$ and Optimization}

The viseme weights to be solved need to satisfy the box constraints within the range $[0,1]$. As we employ the optimization framework of HiFi3DFace \cite{bao2021high}, which uses an iterative optimizer to update parameters, we convert the box constraints to a value range loss $L_{range}$ and truncate the final results after optimization to fulfill the box constraints. The value range loss $L_{range}$ is defined as
\begin{equation}
    L_{range}(\mathcal{X}^{j}) = \frac{1}{ | \mathcal{I}_{upper} | } \sum_{i \in \mathcal{I}_{upper}} (x_i - 1)^2  +  \frac{1}{ | \mathcal{I}_{lower} | } \sum_{i \in \mathcal{I}_{lower}} (0 - x_i)^2 ,
\end{equation}
where $\mathcal{I}_{upper}$ is the set of viseme indices whose parameter $x_i > 1$ and $\mathcal{I}_{lower}$ is the set of viseme indices whose parameter $x_i < 0$ in current iteration. Note that the two sets $\mathcal{I}_{upper}$ and $\mathcal{I}_{lower}$ are re-computed in each iteration with currently estimated parameters.

\subsection{Results and Evaluation}\label{sec.evaltracking}

\subsubsection*{Implementation Details} 

The camera poses and texture parameters are estimated along with the viseme weights, similar to HiFi3DFace \cite{bao2021high}. The weights in Eq. \eqref{eq.totaloptloss} is set to $w_1=0.8$, $w_2=1.0$, $w_3=800$, $w_4=150$, $w_5=1.0$, $w_6=300$, $w_7=100$. Note that the weights for loss terms that are directly applied on viseme weights are two orders of magnitude larger than the other terms, as the average values of the viseme weights are typically much smaller than $1$. The rendered image size from the differentiable renderer is $1280 \times 720$. The procedurally generated weights in Section \ref{sec.procvism} are used to initialize the viseme weights. We use Adam optimizer \cite{kingma2014adam} in Tensorflow to run 250 iterations for each frame, with a learning rate $0.1$ decaying exponentially in every 10 iterations. For each video clip (1 utterance), we first process through the frames sequentially and then process the frames backward in reverse order for a second pass. 





\subsubsection*{Tracking Accuracy and Curve Dynamics}

\begin{figure}[t]
    \captionsetup[subfigure]{aboveskip=-1pt,belowskip=-1pt}
     \centering
     \begin{subfigure}[b]{0.234\textwidth}
         \centering
         \includegraphics[width=\textwidth]{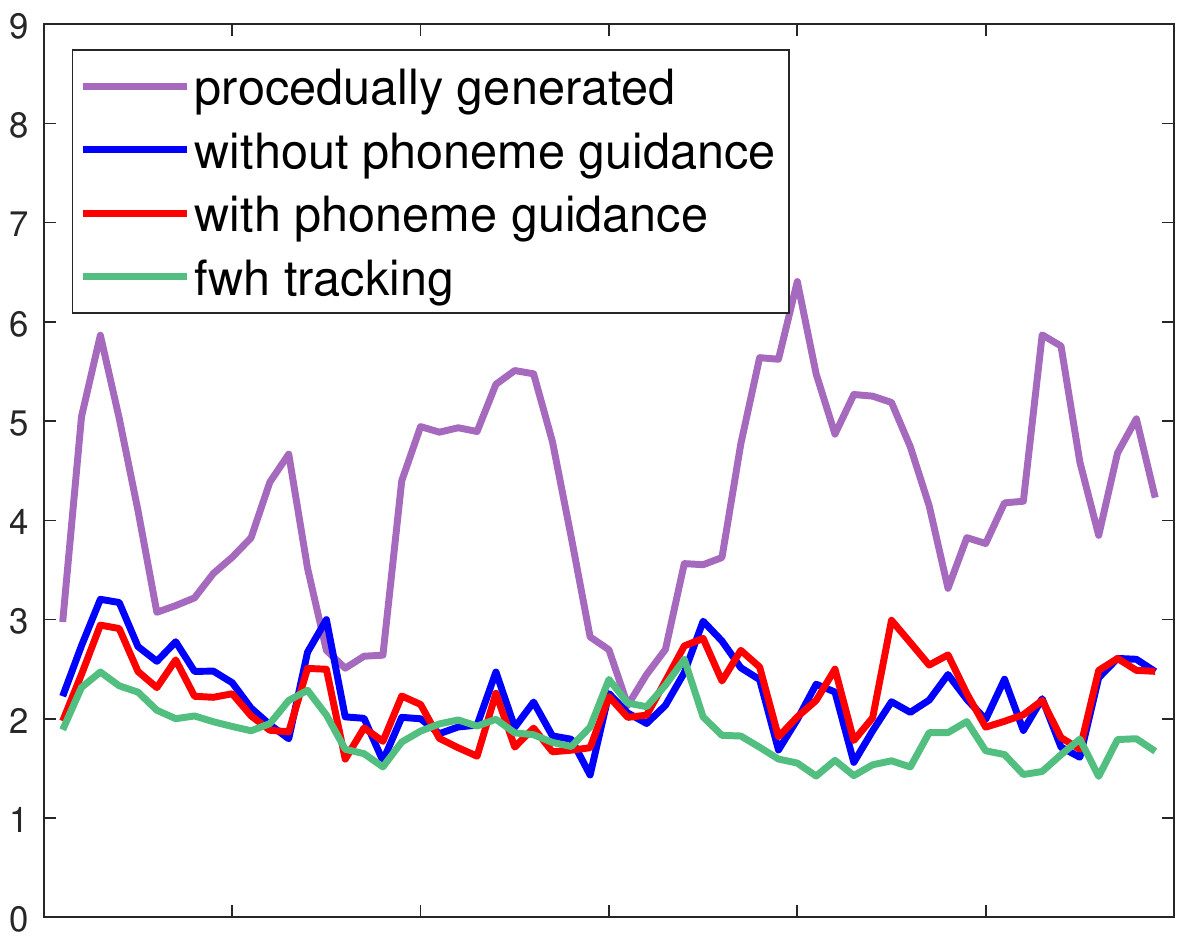}
         \caption{Tracking accuracy}
     \end{subfigure}
     \hfill
     \begin{subfigure}[b]{0.234\textwidth}
         \centering
         \includegraphics[width=\textwidth]{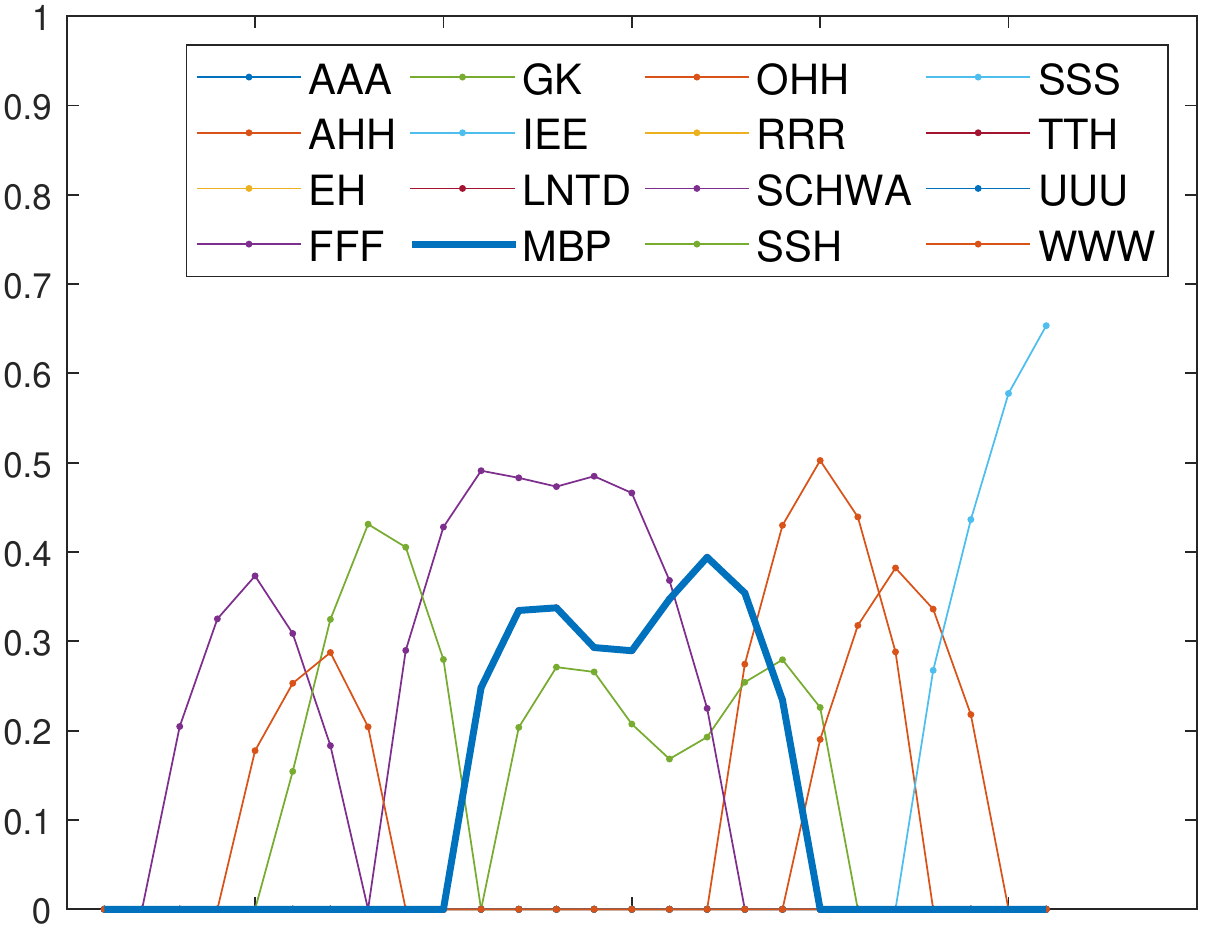}
         \caption{Procedurally generated curves}
     \end{subfigure}
     \hfill
     \begin{subfigure}[b]{0.234\textwidth}
         \centering
         \includegraphics[width=\textwidth]{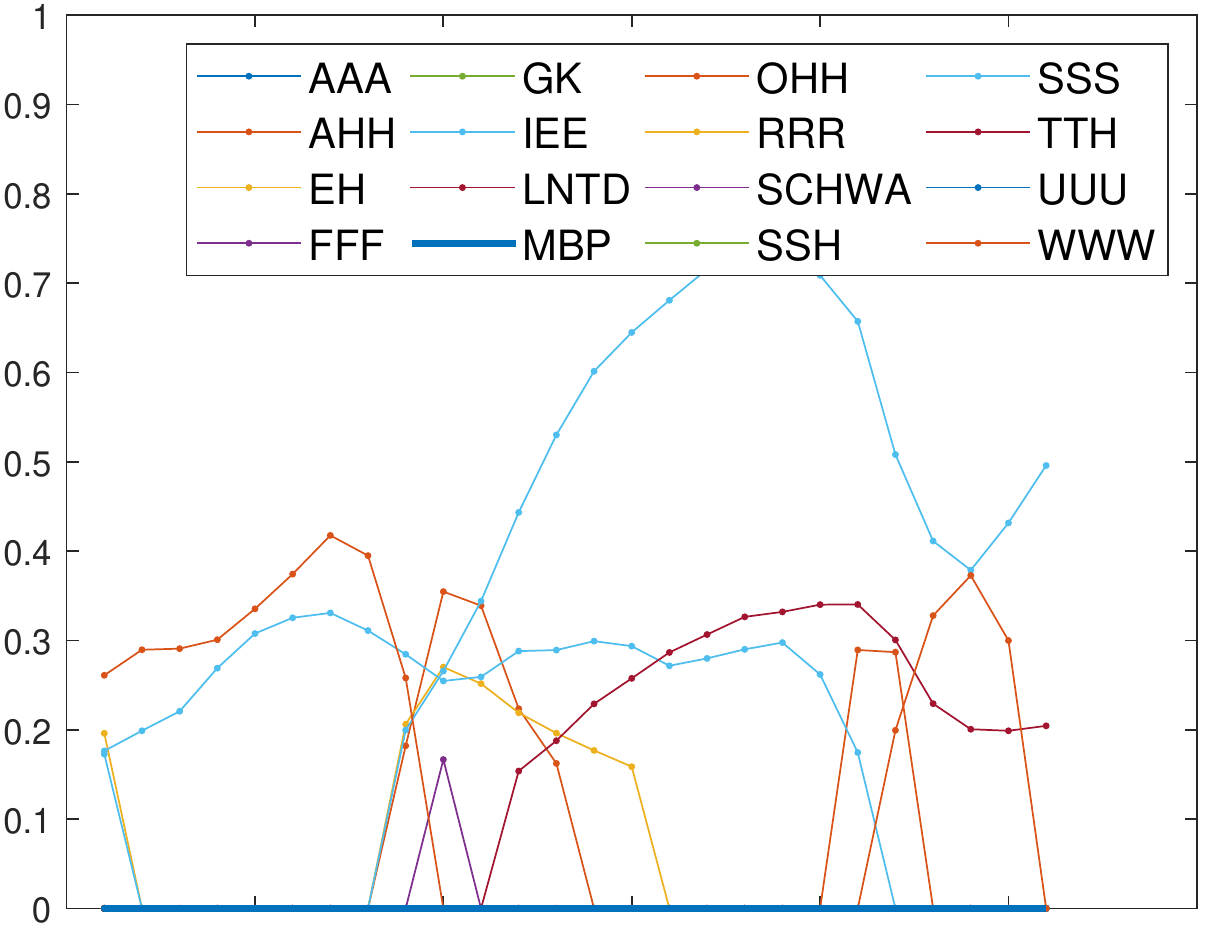}
         \caption{Tracking w/o phoneme guidance}
     \end{subfigure}
     \hfill
     \begin{subfigure}[b]{0.234\textwidth}
         \centering
         \includegraphics[width=\textwidth]{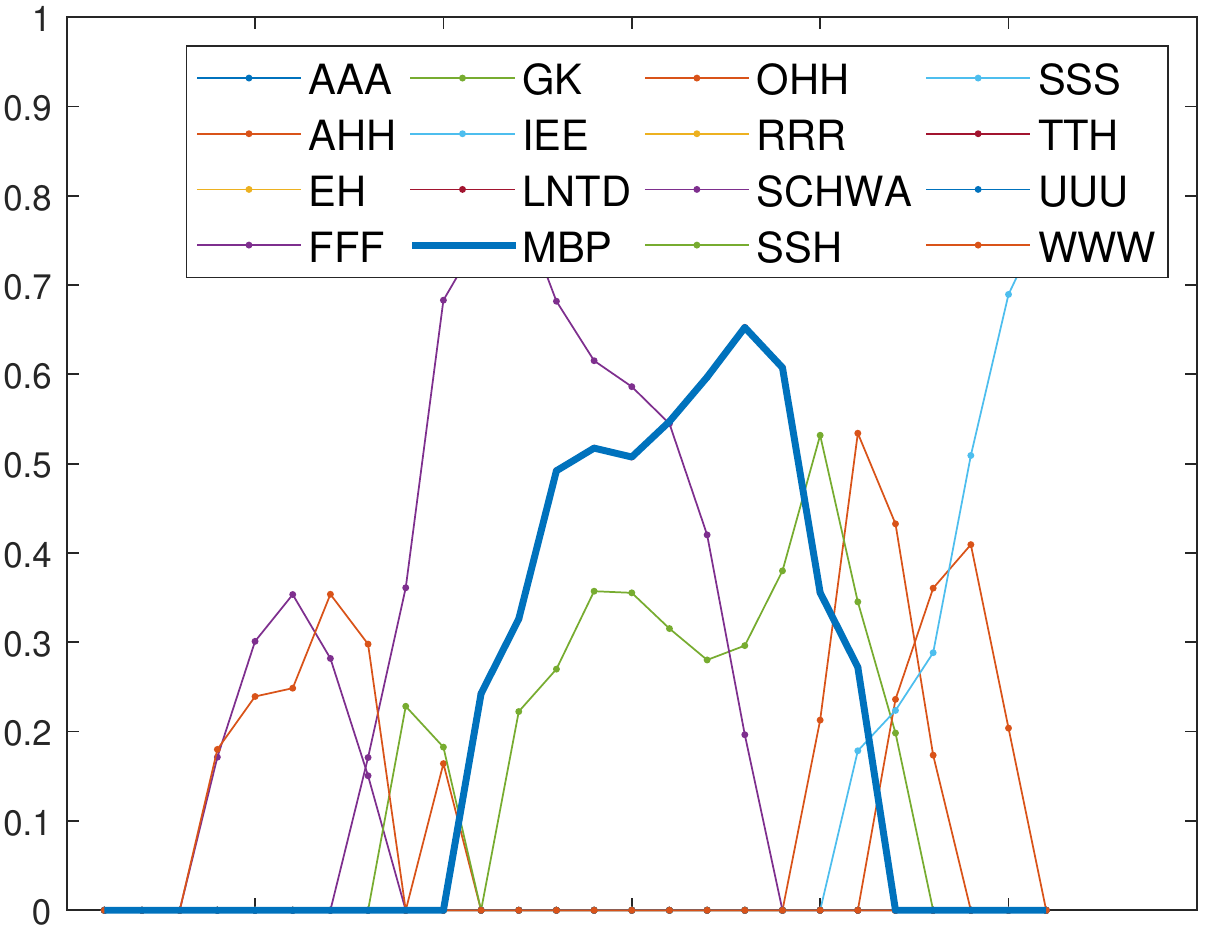}
         \caption{Tracking w/ phoneme guidance}
     \end{subfigure}
  \vspace{-2mm}
  \caption{Comparison of tracking accuracy and viseme curves. Top-left: the comparison of average errors of the facial keypoints around mouth region. The other three plots are viseme curves acquired with different versions of our algorithm. See the text in Section \ref{sec.evaltracking} for detailed explanations.}
  \Description{Comparison of tracking accuracy and viseme curves.}
        \label{fig.vismcomp}
        \vspace{-2mm}
\end{figure}

We first evaluate the tracking accuracy of our viseme curve extraction algorithm. The top-left plot of Figure \ref{fig.vismcomp} shows an example of the comparison of average mouth keypoints errors. Our tracking algorithm using viseme blendshapes, either with or without phoneme guidance, achieves comparable accuracy to the tracking algorithm using a bilinear face model namely FacewareHouse (``fwh'' in short) \cite{cao2014facewarehouse}. While the phoneme guidance in our algorithm does not affect the tracking accuracy, it leads to quite different viseme curves as shown in the bottom two plots in Figure \ref{fig.vismcomp}. Note that although the viseme curves obtained with phoneme guidance (bottom-right) appear similar to the procedurally generated viseme curves (top-right), the tracking accuracy gets largely improved (in top-left plot). It also could be observed from the curves that our tracked viseme curves have less regular shapes with more detailed variations, which better resembles real motion dynamics and would reduce the sense of artificial composition in the resulting animations. 





\begin{figure}[t]
  \centering
  \includegraphics[width=0.99\linewidth]{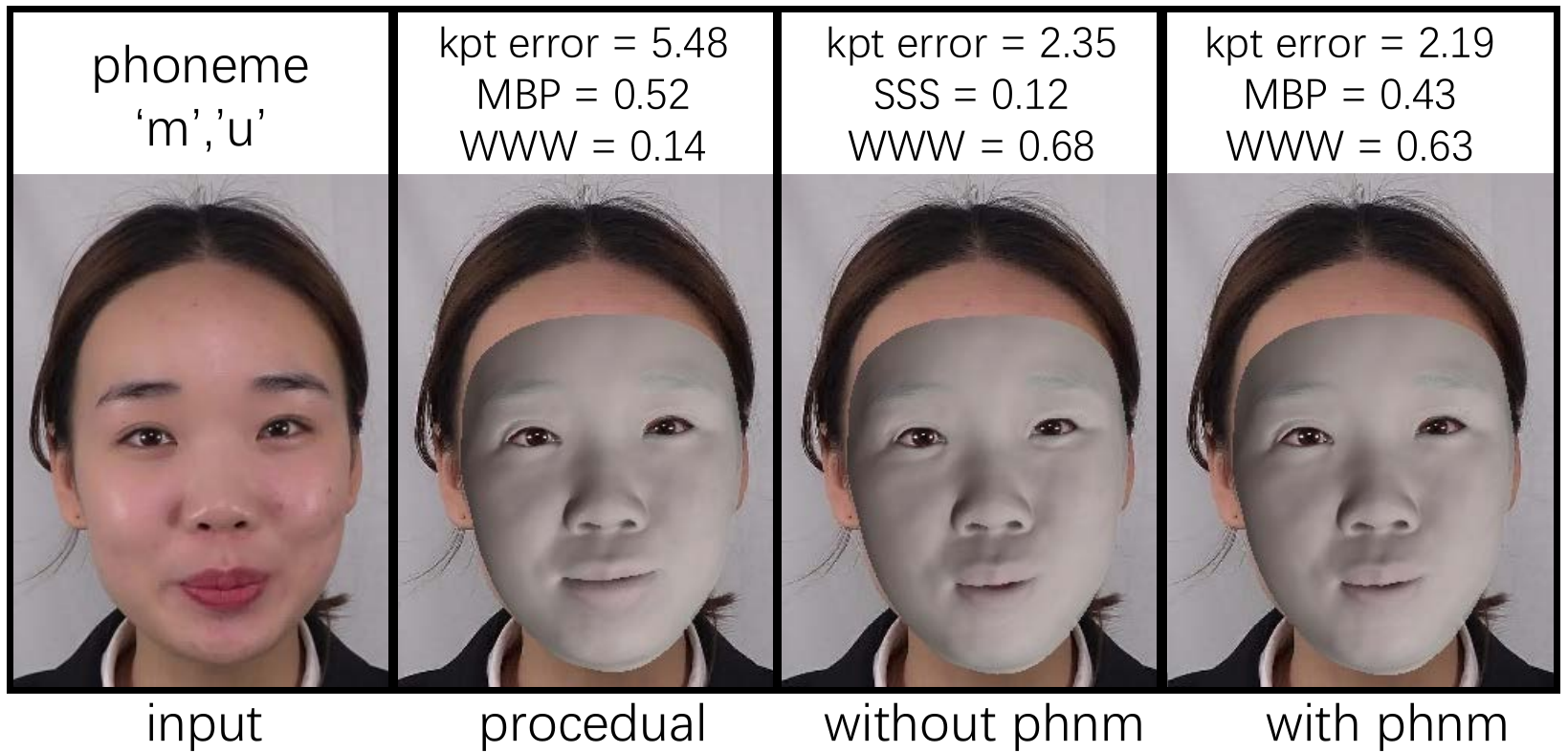}
  \vspace{-2mm}
  \caption{The effects of phoneme guidance in viseme weights extraction from videos. The keypoint errors and the top two activated viseme weights are shown. Our tracked result with phoneme guidance has the lowest error and correctly activates the bilabial phoneme ``MBP''.}
  \Description{The effects of phoneme guidance in viseme weights extraction.}
  \label{fig.vismcompimgs}
  \vspace{-3mm}
\end{figure}


\subsubsection*{Effects of Phoneme Guidance}

We further study the effects of the phoneme guidance to the extracted viseme curves. Comparing the bottom two plots in Figure \ref{fig.vismcomp}, we observe that in some frames, the activated visemes are rather different from the results obtained with and without phoneme guidance. Figure \ref{fig.vismcompimgs} shows an example of such a frame. As the phoneme label of the frame is the transition from `m' to `u', the procedural generation rule mainly activates viseme ``MBP'' and ``WWW''. The resulting keypoints error is large, since the generation rule completely ignores the image information. On the other hand, although the keypoints error of the tracking without phoneme guidance is much smaller, it mistakenly activates viseme ``SSS'' instead of ``MBP''. With the guidance of phonemes, our tracking algorithm activates the correct visemes and achieves satisfying keypoints error. Note that although the differences between activating correct and wrong visemes are unnoticeable in the original tracked video, it would be very different when applied to new characters. For example, the viseme ``MBP'' is the pronunciation action of \textit{bilabial} phonemes, which requires a preparation action that both the upper and lower lips squeeze each other for stopping and then releasing an air puff from mouth. This is rather different from the viseme ``SSS'', which is the pronunciation action of \textit{sibilant} phonemes, made by narrowing the jaw and directing air through the teeth. The differences between them might be difficult to observe solely from images, but they will be carefully distinguished by experienced artists, especially for animations that need artistic exaggerations \cite{osipa2010stop}.




\section{Learning Audio-to-Curves Mapping}\label{sec.mapping}

We extract the viseme curves from all the videos in our dataset ($12,000$ utterances), and split the dataset into a training set ($10,000$ utterances) and a validation set ($2,000$ utterances). 
The goal of the learning is to map each audio track $\mathbf{U}_0 \in \mathbb{R}^{T_U}$ to a sequence of viseme weights $\mathbf{Y} \in \mathbb{R}^{T\times 16}$, which should be close to the extracted viseme weights from videos $\mathbf{X} \in \mathbb{R}^{T\times 16}$ (\textit{i.e.}, $\{\mathcal{X}^j\}_{j=1}^T$). 
%
A major issue with our dataset is that all of the utterances are from the same person with a single tone. 
%
Thus adopting audio features that are pretrained using extra data is vital for the learned mapping to be generalizable to unseen speakers.
%
We extensively investigated various audio features and find Wav2Vec2 \cite{baevski2020wav2vec} pretrained on a large-scale corpus in 53 languages \cite{conneau2020unsupervised} works the best in our setting. 
Moreover, we also studied various sequence-to-sequence mapping backbones including temporal convolutional networks (TCNs), transformers, LSTMs, etc., and observe that Bidirectional LSTM achieves the best performances. 
%
%
We present the best-performing model in Section \ref{sec.mapmodel} and the experimental studies in Section \ref{sec.mapping_result}.

\subsection{Mapping Model}\label{sec.mapmodel}

\subsubsection*{Audio Encoder}  
We denote the network structure of Wav2Vec2 as $\{\phi, \varphi\}$, where $\phi$ stands for the collection of $n_1=7$ layers of Temporal Convolutional Networks (TCN) and $\varphi$ for the collection of $n_2=24$ layers of multi-head self-attention layers which transforms the audio features into contextualized speech representations $\mathbf{U}_1\in\mathbb{R}^{T_W \times D_W}$, \textit{i.e.}, $\mathbf{U}_1 = \varphi(\phi(\mathbf{U}_0))$.
We follow the general finetune strategy for Wav2Vec2, \textit{i.e.}, fixing the pretrained weights for $\phi$ and allowing finetuning for the weights of $\varphi$. 
We next add a randomly initialized linear layer $f_{proj}$ to project the feature dimension from $D_W=1024$ to $D=512$ and an interpolation layer $f_{interp}$ to align the frame rates of the features with viseme curves. Thus, the audio feature can be expressed as $\mathbf{U}_2=f_{interp}(f_{proj}(\mathbf{U}_1)) \in \mathbb{R}^{T\times D}$.
Similar to FaceFormer \cite{fan2022faceformer}, we also employ a periodic positional encoding (PPE) to enhance the audio features, that is $\mathbf{U}_3=\mathbf{U}_2 + \mathbf{P}$, where $\mathbf{P}$ is the PPE.

\subsubsection*{Viseme Curves Decoder}
We employ a one-layer Bidirectional LSTM followed by a fully connected layer to map the audio features to the 16-dimensional viseme weights, that is $\mathbf{Y}=f_{FC}(f_{BLSTM}(\mathbf{U}_3))$.
The model is trained using the L1 loss between the tracked viseme weights $\mathbf{X}$ and the predicted viseme weights $\mathbf{Y}$. 

\subsubsection*{Training} 
SpecAugment~\cite{park2019specaugment} is employed during the training phase. 
We use the Adam optimizer with a fixed learning rate of $10^{-5}$ to optimize a single utterance in each training iteration. The model is trained for 100 epochs on the training set. 



\begin{table}[t!]
  \centering
  \caption{Ablation study for different audio features. 
  The listed values are L1 reconstruction errors multiplied by $10^{2}$.
  \textit{Volume} stands for the generality test performed on the validation set with augmented volume of the test audio. The same goes for \textit{pitch}, \textit{speed} and \textit{noise}.
  }
  \vspace{-2mm}
  \resizebox{\linewidth}{!}{
    \begin{tabular}{l|c|cccc}
    \hline
    \hline
    \multicolumn{1}{c|}{{\diagbox{Feature}{Metric}}} & \multicolumn{1}{c|}{validation} & \multicolumn{1}{c}{volume} & \multicolumn{1}{c}{pitch} & \multicolumn{1}{c}{speed} & \multicolumn{1}{c}{noise} \\
          \hline
    FBank & 3.21 & 3.32 & 4.59  & 3.39  & 4.42  \\
    LPC & 5.26 & 5.27 & 6.50  & 5.35  & 5.93  \\
    PPG \cite{huang2021speaker} & 3.01 & 3.01 & 3.07  & 7.50  & 3.13  \\\hline
    Wav2Vec2$^1$ \cite{baevski2020wav2vec} & 1.88   & 1.89   & 2.48  & 2.29     & 2.47  \\
    Wav2Vec2$^2$ \cite{baevski2020wav2vec} & 1.82 & 1.83 & 2.55 & 2.19  &  2.27  \\
    Wav2Vec2$^3$ (ours) \cite{conneau2020unsupervised} & \textbf{1.77} & \textbf{1.77} & \textbf{2.44} & \textbf{2.06} & \textbf{2.09} \\
    \hline
    \hline
    \end{tabular}%
    }
  \label{table.feature}%
  \vspace{-0.2cm}
\end{table}%

\begin{table}[t!]
  \centering
  \caption{Ablation study for different decoder backbones. 
  The listed values are L1 reconstruction errors multiplied by $10^{2}$.
  }
  \vspace{-2mm}
  \resizebox{\linewidth}{!}{
    \begin{tabular}{l|c|cccc}
    \hline
    \hline
    \multicolumn{1}{c|}{{\diagbox{Decoder}{Metric}}} & \multicolumn{1}{c|}{validation} & \multicolumn{1}{c}{volume} & \multicolumn{1}{c}{pitch} & \multicolumn{1}{c}{speed} & \multicolumn{1}{c}{noise} \\
          \hline
    TCN & 1.77 & 1.78 & 2.51  & 2.11  & 2.11  \\
    LSTM & 1.77 & 1.78 & 2.50  & 2.12  & 2.16  \\
    Transformer & 1.79   & 1.80    & 2.54     & 2.13 & 2.14  \\\hline
    BLSTM (ours) & \textbf{1.77} & \textbf{1.77} & \textbf{2.44} & \textbf{2.06} & \textbf{2.09} \\
    \hline
    \hline
    \end{tabular}%
    }
  \label{table.decoder}%
  \vspace{-0.2cm}
\end{table}%

\subsection{Results and Evaluation}\label{sec.mapping_result}

\subsubsection*{Robustness Evaluation} 
We evaluate the reconstruction loss of the predicted viseme weights via L1 measurement on the validation set. Moreover, in order to evaluate the algorithm generality for audio with various pitches, volumes, speeds, and noises, we augment the validation set by randomly adjusting the pitch in the range of -3 semitone to 3 semitones, volume in the range of -10 dB to +10 dB, speed in the range of 0.7 to 1.3, and add additional Gaussian noise with amplitude smaller than 0.005.

\subsubsection*{Ablation Study on Audio Features} 
We investigate a variety of audio features, including filterbank (FBank), linear predictive coding (LPC), Phonetic Posteriorgrams (PPG), and Wav2Vec2 trained on different datasets and network structures, as shown in Table~\ref{table.feature}. 
Wav2vec2$^1$ denotes the base model \cite{baevski2020wav2vec} ($n_1=7$ ,$n_2=12$) pretrained on Librspeech dataset ~\cite{panayotov2015librispeech} which contains 960 hours of English speaking audio. It is also the pretrained model adopted in FaceFormer~\cite{fan2022faceformer}; Wav2Vec2$^2$ is pretrained on the same dataset with a larger model ($n_1=7$, $n_2=24$); and Wav2Vec2$^3$ is the model we utilize in this paper, which is pretrained on datasets containing 53 languages. The PPG features with dimension $\mathbb{R}^{T \times 128}$ are produced using the same process as Huang et al.~\shortcite{huang2021speaker}.
Compared with handcrafted features like FBank and LPC, the self-supervised pre-trained deep feature extracted from Wav2Vec2 can substantially increase the performance. 
Specifically, comparing the results of Wav2Vec2$^2$ and Wav2Vec2$^3$, one can observe that pretraining on multilingual datasets can further boost the performance of this task.

\subsubsection*{Ablation Study on Decoder Backbones} 
To fairly compare different decoder backbones, we keep the same number of layers and hidden states dimension for each compared decoder, \textit{i.e.} the number of layers $l=1$ and the feature dimension $D=512$. Results in Table~\ref{table.decoder} show that any decoder equipped with the Wav2Vec2 audio features are sufficient to produce high-quality viseme curve output. In the meantime, BLSTM is slightly better than other decoders due to its effectiveness in aggregating neighborhood information. Note that we refer to Transformer's encoder network structure in this ablation study, which utilizes multi-head self-attention layers for feature extraction. FaceFormer~\cite{fan2022faceformer}, in contrast, adopts the Transformer-based decoder, which is an autoregressive model with a mask for cross-modal alignment. 



\subsubsection*{Size of Training Dataset} 
\begin{figure}[t!]
  \centering
  \includegraphics[width=0.96\linewidth]{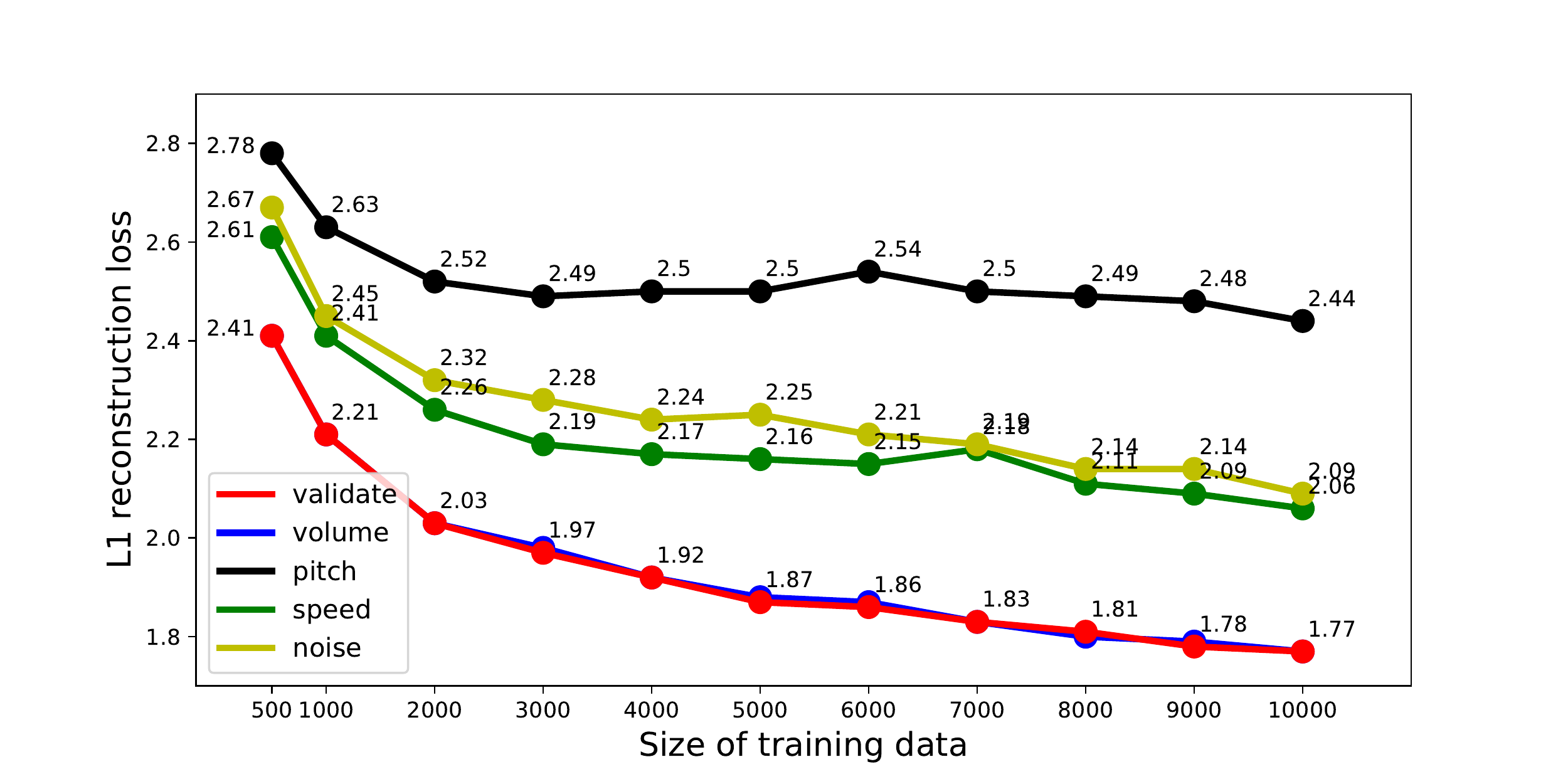}
  \vspace{-2mm}
  \caption{L1 reconstruction losses on validation set for our model trained on datasets with various sizes. All experiments are trained for $10^{6}$ iterations.}
  \label{fig.datasize}
\end{figure}

We further show the performances of our model when trained with different dataset sizes in Figure ~\ref{fig.datasize}.
The records show that, with full data our model achieves the best performances, while with fewer data like 3000 utterances, the model yields reasonably good results and could achieve a good balance between data cost and performance. 

\begin{table}[t]
  \centering
  \caption{Comparison with state-of-the-art deep models. 
  The listed values are L1 reconstruction errors multiplied by $10^{2}$.
  }
  \vspace{-2mm}
  \resizebox{\linewidth}{!}{
    \begin{tabular}{l|c|cccc}
    \hline
    \hline
    \multicolumn{1}{c|}{{\diagbox{Method}{Metric}}} & \multicolumn{1}{c|}{validation} & \multicolumn{1}{c}{volume} & \multicolumn{1}{c}{pitch} & \multicolumn{1}{c}{speed} & \multicolumn{1}{c}{noise} \\
          \hline
    Karras et al. \shortcite{karras2017audio} & 4.11 & 4.15 & 5.79  & 7.42  & 6.34  \\
    Huang et al. \shortcite{huang2021speaker} & 2.38 & 2.38 & 2.45  & 7.43  & 2.69  \\
    FaceFormer \cite{fan2022faceformer} & 2.10  & 2.10  & 2.69  & 2.51  & 2.89  \\
    \hline
    Ours &  \textbf{1.77} & \textbf{1.77} & \textbf{2.44} & \textbf{2.06} & \textbf{2.09}     \\
    \hline
    \hline
    \end{tabular}%
    }
  \label{table.sota}%
  \vspace{-0.3cm}
\end{table}%


\subsubsection*{Comparison with State-of-the-arts} 
We adopt three recent deep learning based audio-driven speech animation models \cite{karras2017audio,huang2021speaker,fan2022faceformer} into our settings and compare their performances with our model. 
Specifically, we modify the last fully connected layer in their networks to predict the 16-dimensional viseme weights, and then train them using our training dataset for 100 epochs. 
The results can be found in Table~\ref{table.sota}. 
Our model achieves the best performances, especially when dealing with different audio variations in terms of pitch and speed. 

\subsubsection*{Generalize to Unseen Speakers and Other Languages} 
Although our training dataset only contains audio from a single speaker and is mostly in Chinese language, our model generalizes well to unseen speakers and other languages, thanks to the powerful pretrained audio feature model. 
We show various animation examples in the supplementary video including that are driven by different speakers in different languages.

\begin{figure}[t]
  \centering
  \includegraphics[width=0.98\linewidth]{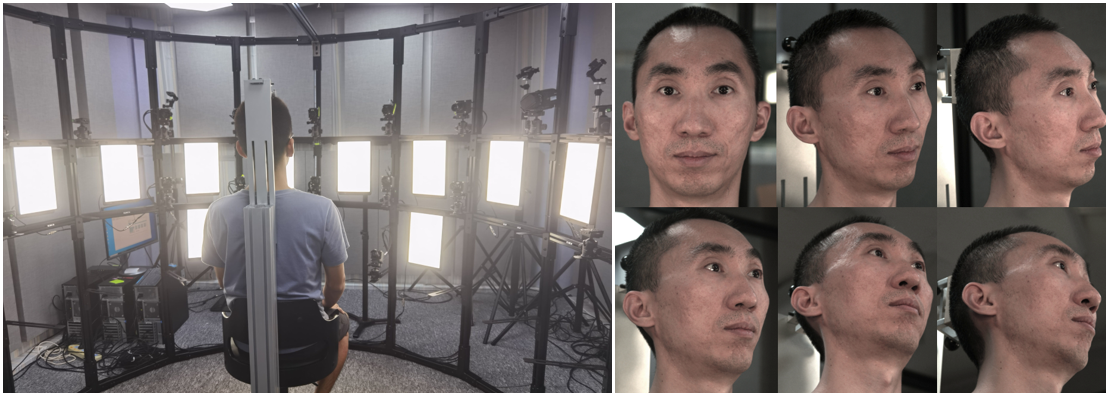}
  \vspace{-2mm}
  \caption{Our simple camera array system consists of 12 synchronized 4K/60fps video cameras and 10 square LED lights. Examples of the acquired images are shown on the right.}
  \Description{Our camera array system and the acquired images.}
  \vspace{-2mm}
  \label{fig.camarraysys}
\end{figure}

\begin{figure}[t]
  \centering
  \includegraphics[width=0.98\linewidth]{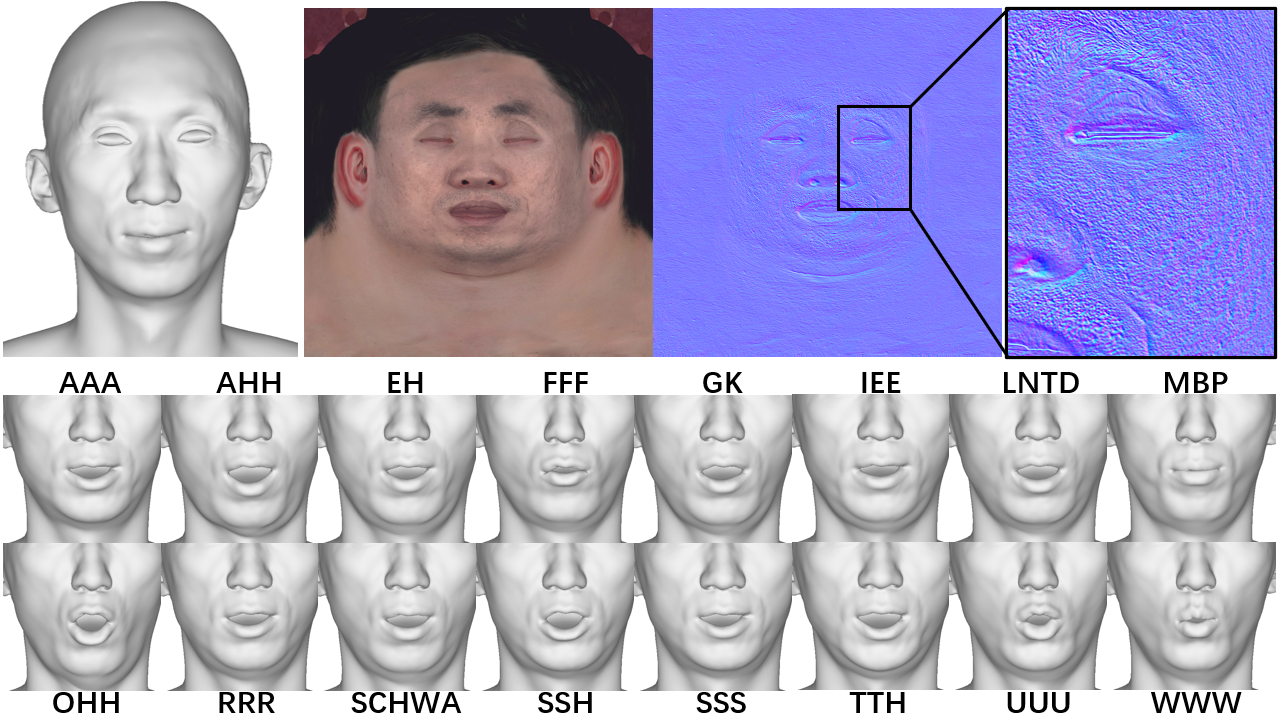}
  \vspace{-2mm}
  \caption{An example of the facial assets and viseme blendshapes acquired by viseme scanning. See the supplementary video for animations.}
  \Description{The facial assets and viseme blendshapes acquired by viseme scanning. See the supplementary video for animations.}
  \vspace{-2mm}
  \label{fig.vismscanresult}
\end{figure}

\section{Speech Animation Production}\label{sec.animproduction}

With the above trained audio-to-curves model, speech animation can be then automatically produced from audio tracks, given properly prepared viseme blendshapes for each character. As each viseme blendshape has a clear mapping to phonemes, the blendshapes preparation for a new character is artist-friendly. However, manually sculpting high-quality viseme blendshapes would still require large amounts of work. In this section, we present an approach to efficiently acquire high-fidelity blendshape assets by scanning the viseme shapes of an actor. The scanned blendshape assets can be further converted to bone-pose assets for bone animations, which are desired in resource-sensitive applications.

\subsection{Viseme Blendshape Scanning}\label{sec.visemescan}

To acquire high-fidelity viseme blendshapes of an actor, we use a simple camera array system (shown in Figure \ref{fig.camarraysys}) to capture multi-view videos of the actor while performing pronunciation actions of the phonemes in Figure \ref{fig.viseme16}. We select the video frames corresponding to the 16 visemes and reconstruct the 3D face models using a multi-view stereo algorithm \cite{openmvs2020}. The 3D face models are then aligned and deformed to a fixed-topology mesh template using a non-rigid registration algorithm \cite{bouaziz2016modern}. Figure \ref{fig.vismscanresult} shows an example of the resulting facial assets and topology-consistent viseme blendshapes. We further extract the texture maps and normal maps using a method similar to Beeler et al. \shortcite{beeler2010high} for each viseme blendshape to get dynamic texture maps \cite{li2020dynamic}. Accessories like eyeballs and mouth interiors are attached for each viseme blendshape using the approach described in Bao et al. \shortcite{bao2021high}. With these facial assets, the viseme weights produced by the audio-to-curves model can be directly applied to get high-quality, realistic facial animations (see the supplementary videos).

\begin{figure}[t]
  \centering
  \vspace{-2mm}
  \includegraphics[width=0.99\linewidth]{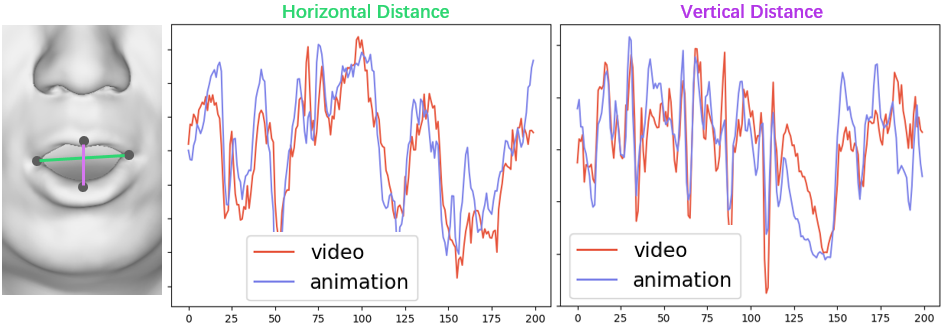}
  \vspace{-2mm}
  \caption{Comparison of lip motion curves between real video and animation. We record the temporal sequences of the horizontal and vertical distances between two pairs of keypoints from both real video and animation. The comparison demonstrates that the lip motions in the produced animation well resemble real lip motions.}
  \Description{Comparison of lip motion curves between real video and animation.}
  \vspace{-3mm}
  \label{fig.baocurves}
\end{figure}

We can further validate the motion accuracy of the animations by comparing the lip motions between the recorded speech video of the actor and the produced animation. Figure \ref{fig.baocurves} shows an example clip of the comparison. The result shows that the lip motions in the produced animation well resemble real lip motions, including rich realistic motion details. Note that the viseme curves applied to the produced animation are learned from data of the actress in our large-corpus dataset, while the real videos are recorded from the new actor. The experiment demonstrates that our learned viseme dynamics have good generalization performances for producing realistic lip animations when applied to new characters.

\begin{figure}[t]
  \centering
  \includegraphics[width=0.99\linewidth]{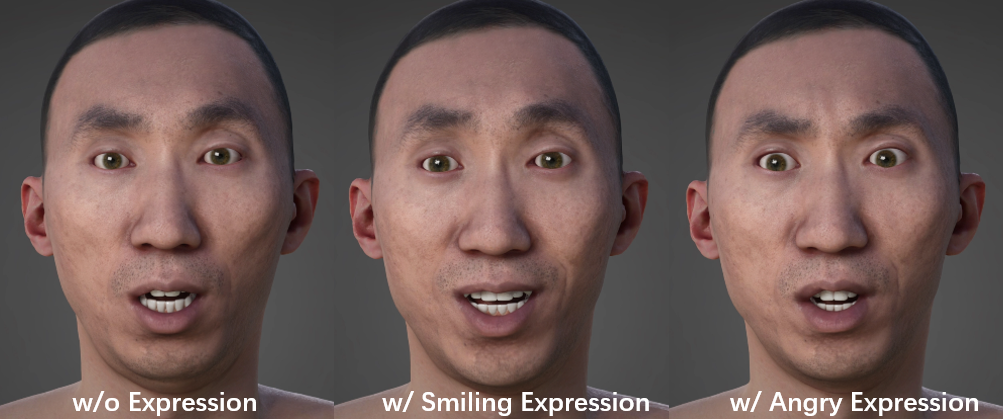}
  \vspace{-2mm}
  \caption{Speech animation with expressions. The viseme blendshapes are scanned from the actor with different expressions, while the viseme curves are the same as the animation without expressions. The videos are provided in the supplementary material.}
  \vspace{-2mm}
  \Description{Speech animation with expressions.}
  \label{fig.baoexpr}
\end{figure}

\begin{figure}[t]
  \centering
  \includegraphics[width=0.99\linewidth]{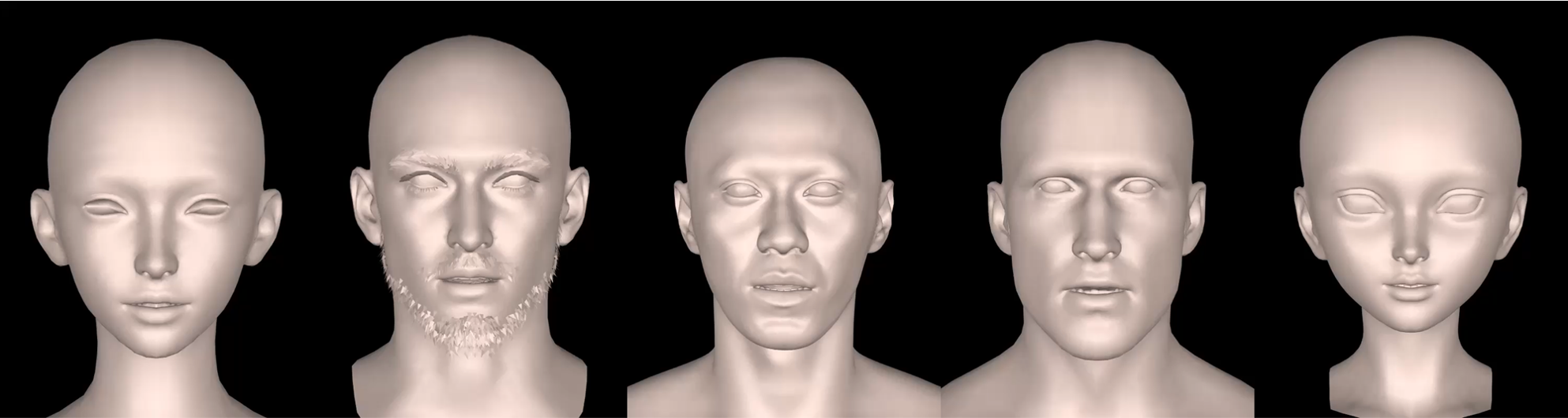}
  \vspace{-2mm}
  \caption{A snapshot of our speech-driven animation results. The characters are driven by the same set of viseme curves predicted by our audio-to-curves model. Videos can be found in the supplementary materials.}
  \vspace{-2mm}
  \Description{A snapshot of our speech-driven animation results.}
  \label{fig.animmultiple}
\end{figure}

The approach can be further applied to produce speech animations with facial expressions. We scan a set of expression visemes of the actor during his phoneme pronunciations with a certain expression, e.g., smiling or angry. Then the same viseme curves as previous results are applied to the expression visemes. Figure \ref{fig.baoexpr} shows a snapshot of the animation with two different expressions. Please refer to the supplementary for the animation videos.

\subsection{Retargeting and Bone Animation}\label{sec.boneanim}

To animate characters that cannot be scanned from actors, we use the deformation transfer algorithm \cite{sumner2004deformation} to generate the viseme blendshapes from artistic templates or scanned characters. In many resource-sensitive applications with real-time graphics engines like Unity or Unreal Engine, bone-based animations are preferred rather than blendshape-based animations. We use the SSDR algorithm \cite{LeDeng2012} to convert each viseme blendshape to a set of bone pose parameters, and then directly apply the predicted viseme curves to blend the bone-pose assets to get bone animations. The pose blending technique is to simply perform linear interpolation on translation/scale parameters and spherical linear interpolation (SLERP) on quaternion-based rotation parameters, which is typically provided by real-time graphics engines, e.g., the Pose Asset Animation \cite{ue4poseassets} in Unreal Engine or the Direct Blend Tree \cite{unity3dblendtree} in Unity Engine.

\subsection{Results}

In the supplementary video, we present visual comparisons of our results to the results reported by other state-of-the-art audio-driven 3D facial animation approaches \cite{karras2017audio,taylor2017deep,cudeiro2019capture,richard2021meshtalk,fan2022faceformer}. 
The lip motions in our results are clearly more realistic and natural. 
Note that quantitative comparisons to other work are not possible due to different animation mechanisms and facial rigs, as also noted in previous work \cite{edwards2016jali,zhou2018visemenet}. 
Besides, we present various speech-driven animations of different characters in the supplementary video. Figure \ref{fig.animmultiple} shows a snapshot of the animation of different characters using the same set of viseme curves predicted by our audio-to-curves model, given personalized viseme blendshapes of each character.

\section{Conclusion}

We have introduced a speech animation approach that can produce realistic lip-synchronized 3D facial animation from audio. Our approach maps input audio to animation curves in a viseme-based parameter space, which is artist-friendly and can be easily applied to new characters. We demonstrated high-quality animations in various styles produced by our approach, including realistic and cartoon-like characters. The proposed approach can serve as an efficient solution for speech animation in many applications like game production or AI-driven digital human applications. 

\subsubsection*{Limitations and Future Work}

Our approach does not capture the motion dynamics of tongues during speaking. 
We associate a static teeth and tongue pose to each viseme blendshape and rely on the viseme curves to yield the animation of teeth and tongue. 
Although in most cases the animation of mouth interiors is unnoticeable due to dark lighting in the mouth, sometimes the unnatural motions would be protruded, especially when the mouth interiors are illuminated. 
We notice a recent work \cite{medina2022speech} addressing this problem and it might deserve more investigation in the future.

\appendix

\vspace{5mm}

\section*{Appendix}

\section{Details of Network Architectures}

\subsection{Proposed Mapping Model}
\begin{figure}[h!]
  \centering
  \includegraphics[width=0.99\linewidth]{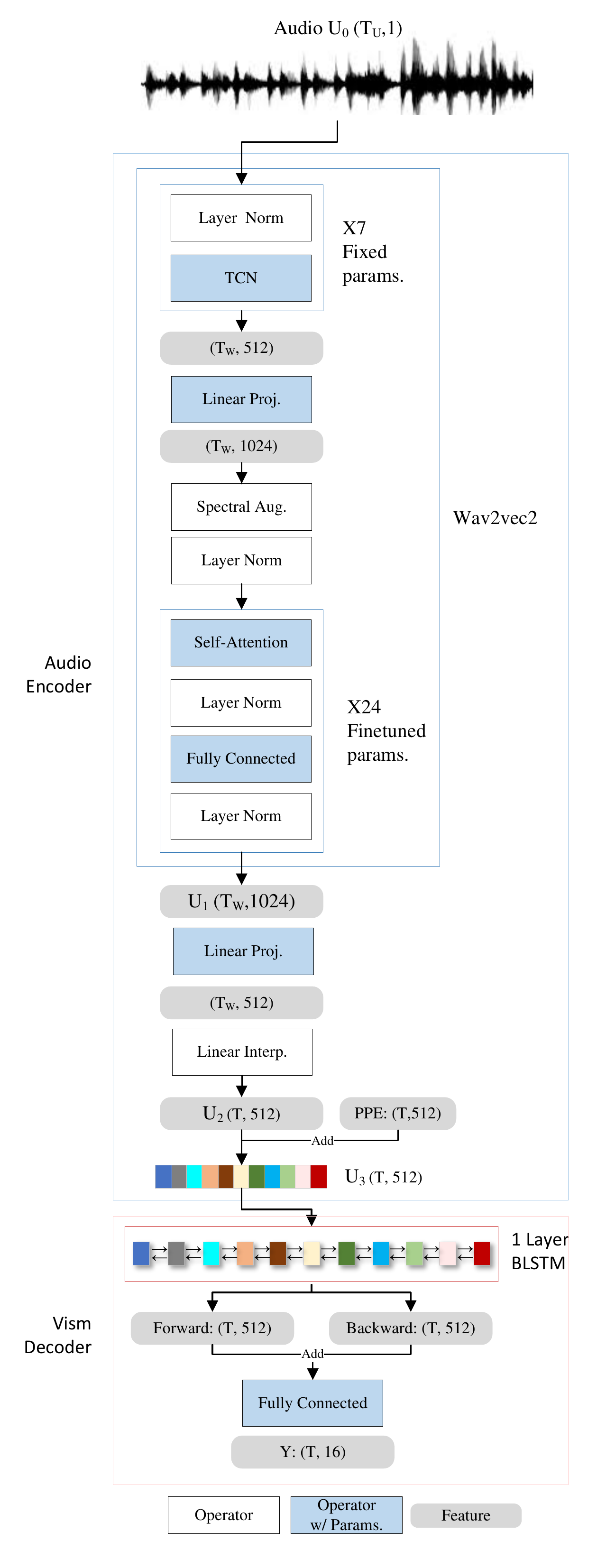}
  \vspace{-6mm}
  \caption{The architecture of our audio-to-curves mapping model.}
  \label{fig.model}
  \vspace{-2mm}
\end{figure}

The network architecture of the proposed mapping model is shown in Fig.~\ref{fig.model}. We omit ReLU activation functions in each TCN and fully connected layer. The network takes the audio $\mathbf{U}_0$ with shape $\mathbb{R}^{T_U}$ as input and output the predicted viseme $\mathbf{Y}\in\mathbb{R}^{T\times 16}$.

\subsection{Ablation Study on Decoder Backbones}
\begin{figure}[h!]
  \centering
  \includegraphics[width=0.75\linewidth]{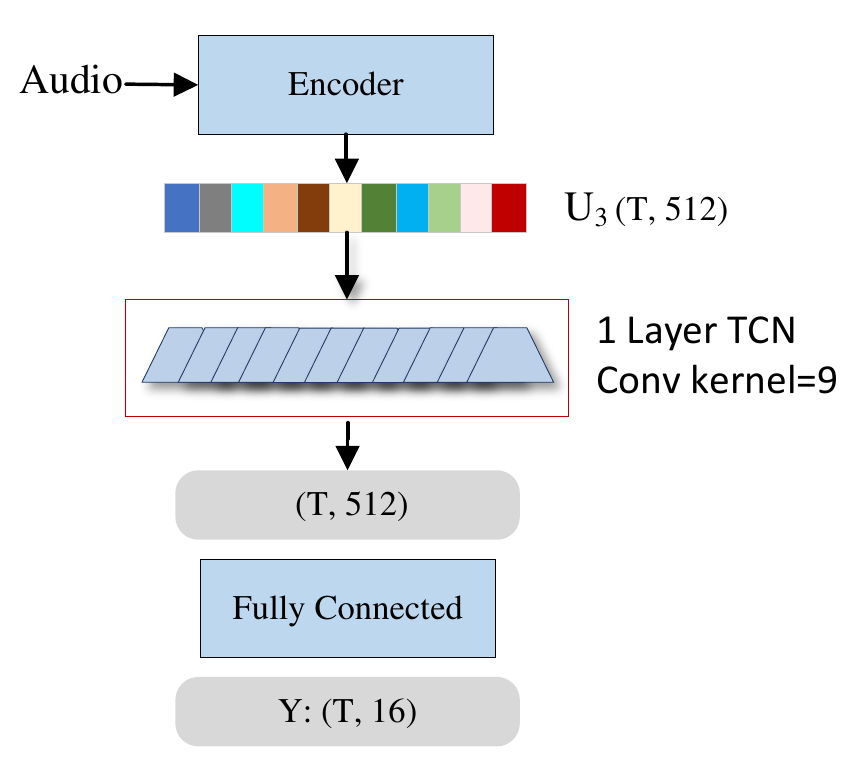}
  \vspace{-2mm}
  \caption{The TCN-based decoder in our ablation study.}
  \label{fig.tcn}
  \vspace{-2mm}
\end{figure}

\begin{figure}[h!]
  \centering
  \includegraphics[width=0.85\linewidth]{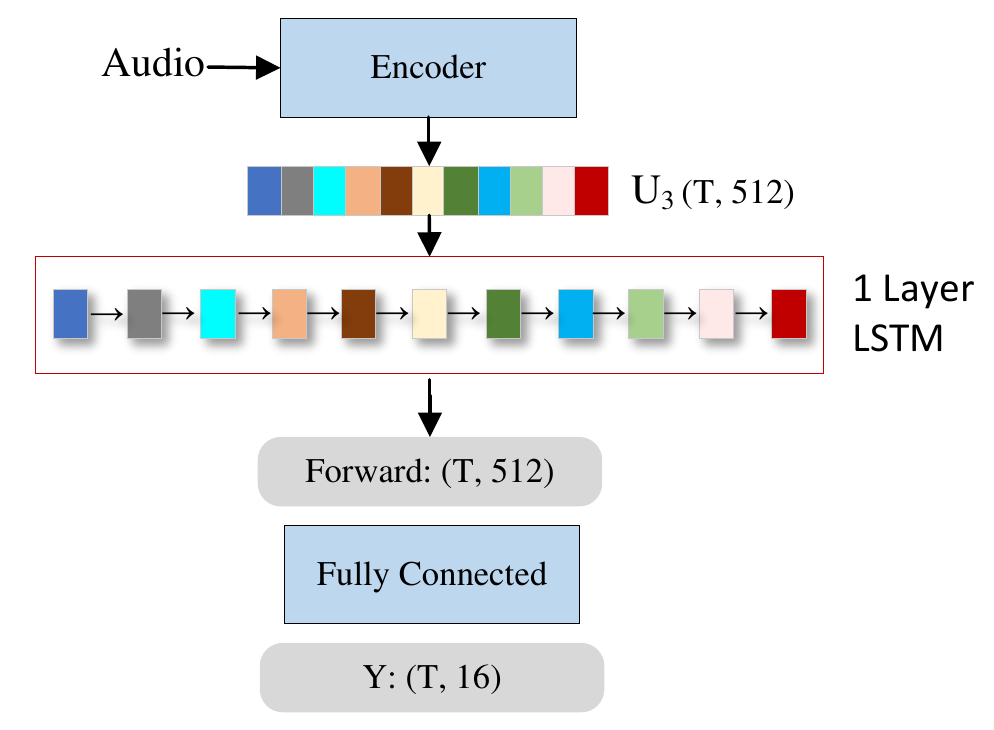}
  \vspace{-2mm}
  \caption{The LSTM-based decoder in our ablation study.}
  \label{fig.lstm}
  \vspace{-2mm}
\end{figure}

\begin{figure}[h!]
  \centering
  \includegraphics[width=0.99\linewidth]{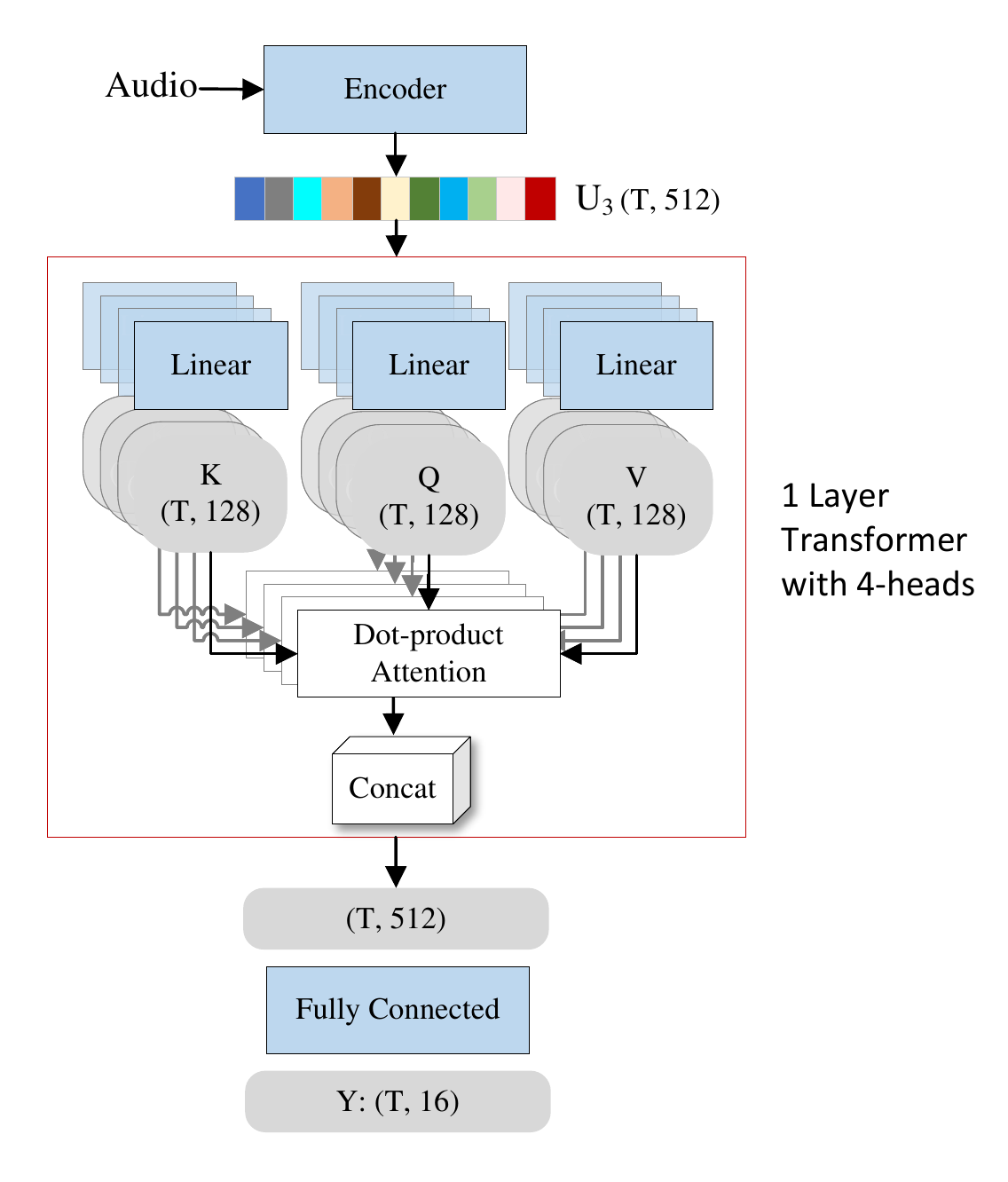}
  \vspace{-2mm}
  \caption{The Transformer-based decoder in our ablation study.}
  \label{fig.transformer}
  \vspace{-2mm}
\end{figure}

In Fig.~\ref{fig.tcn}, ~\ref{fig.lstm} and ~\ref{fig.transformer}, we also show the detailed implementation for the decoders we compared in our ablation study. Note that in this study, we adopt the multi-head self-attention mechanism as the transformer-based decoder.

\clearpage

\bibliographystyle{ACM-Reference-Format}
\bibliography{3dfacebib}

\clearpage









\end{document}